# Consistent Static Models of Local Thermospheric Composition Profiles

J. M. Picone[1,3], J. T. Emmert[2], D.P. Drob[2]

[1]Department of Physics and Astronomy, George Mason University, Fairfax VA 22030
[2]Space Science Division, Naval Research Laboratory, Washington, DC 20375
[3]Emeritus, Naval Research Laboratory, Washington, DC 20375

## Abstract

The authors investigate the ideal, nondriven multifluid equations of motion to identify consistent (i.e., truly stationary), mechanically static models for composition profiles within the thermosphere. These physically faithful functions are necessary to define the parametric core of future empirical atmospheric models and climatologies. Based on the strength of interspecies coupling, the thermosphere has three altitude regions: (1) the lower thermosphere (herein $z < \sim100$ km), in which all species move together at the composite fluid velocity with an effective particle mass equal to the average particle mass of the composite fluid; (2) the upper thermosphere (herein $z > \sim200$ km), in which the species flows are approximately uncoupled; and (3) a transition region in between, where the effective species particle mass and the effective species vertical flow interpolate between the solutions for the upper and lower thermosphere. We place this view in the context of current terminology within the community, i.e., a "fully−mixed" (lower) region and an upper region in "diffusive equilibrium (DE)." The latter condition, DE, currently used in empirical composition models, does not represent a truly static composition profile in the presence of finite thermal diffusion. Rather, species−by−species hydrostatic balance is a consistent (i.e., stationary) static representation of vertical thermospheric composition profiles.

## 1. Introduction

### 1.1. Background

A physically faithful representation of the core thermospheric composition profile is essential for an empirical model to fill gaps in the extant database and to extract from the data thermospheric variability on daily and longer time scales [Picone et al., 2013]. To fulfill this requirement, the authors describe a class of static profile solutions to the governing equations of motion. These solutions will serve as the core of the next-generation Mass Spectrometer Incoherent Scatter Radar (MSIS®) empircial atmospheric model (e.g., Picone et al. [2002]).

In particular, the paper characterizes static altitude profiles of thermospheric composition and develops a physically consistent, static, one-dimensional (1D) thermospheric model of composition, excluding external drivers (e.g., geomagnetic or solar), nonstatic boundary conditions, and chemical and photochemical processes. Such models serve as the core representation of empirical models of the thermosphere, as described below.  Judicious selection of a parametric temperature profile renders these static composition models integrable in closed form, a distinct advantage for climatological or empirical representations. Present empirical models, e.g., Hedin [1987], assume a Bates temperature profile [Bates, 1959] above ~ 100 km, the physics of which has been discussed by Chamberlain and Hunten [1987] and recently by Picone et al. [2013]. Appendix A provides an example of a closed-form model using the Bates temperature profile but does not include a specific extension of the temperature profile through the lower thermosphere. For the latter, Hedin [1987] has shown that a polynomial approximation to 1/T renders the static composition model integrable in closed form and appears to be sufficiently flexible to fit new data sets.

### 1.2. Terminology

Here the term "static" connotes a limit or solution of the governing fluid equations in which all species velocities and their total time derivatives (i.e., forces) are zero (e.g., Goldstein, [1950, p. 15]. Further, Goldstein [p. 318] assures us that a system in a static state, as defined here, will remain so over time (has a time derivative of zero), i.e., that the static state is "stationary." To avoid questions of rigor in this assurance, we add to our definition the explicit constraint that a static model be stationary.

A complication in thermospheric physics is that a model can have an instantaneous or initial velocity field that is zero and therefore could be interpreted as static while failing to meet the constraint of stationarity. In addition, the term "stationary" bears such close resemblance to "static," yet finds application well beyond the realm of static mechanical states. For these reasons, we sometimes use the term "consistent" to imply that a given model has an instantaneous velocity of zero (is apparently static) and that the velocity remains zero over time. This is relevant when discussing so called "diffusive equilibrium," which is generally not static and is therefore not a consistent static state.

Unless unavoidable, this paper uses terms like "force balance" rather than "(mechanical) equilibrium" or concepts associated with mechanical equilibrium, even though the above definition of a static model fits the definition of an equilibrium fluid state by at least some authors (e.g., Dutton [1986], p.195; Goldstein [1950], p.318). Unfortunately, consideration of equilibrium extends the discussion to topics that are irrelevant to this paper: As mentioned in Section 4.1, a major affiliated topic is the response of a system to a small perturbation of a given initial state or position. Because the species density, momentum, and energy fields are coupled via the governing conservation equations, such a response includes changes in the velocity field, even from an initial velocity field of zero. (Static models exclude a nonzero velocity.) Depending on the growth of the perturbation, one then characterizes a mechanical equilibrium state as "stable," "unstable," or "neutral" [e.g., Goodman and Warner, 1964].

The present paper does not involve perturbations to a given static fluid state; rather, the key issue regards the longevity of a particular (unperturbed) composition profile. A truly static, physics-based profile can serve as a core empirical representation, of which the defining parameters may be expanded in harmonic and polynomial series to represent time scales and physical dependences to be extracted from thermospheric data (e.g., Picone et al. [2013]). Present empirical models of thermospheric composition and temperature do not include a coupled velocity model (next section). As the mechanical stability of a fluid state necessarily includes dynamics (nonzero velocity), topics related to mechanical equilibrium are not relevant to our target application.

Finally, this investigation leads naturally to the identification of three thermospheric regions, which are called the "lower thermosphere ($z < \sim100$ km)," the "upper thermosphere ($z < \sim200$)," and an intervening transition region. Our boundaries will differ somewhat from the standard definitions within the space science community. Even the standard definitions are not universal. For example, the term "lower thermosphere" often refers either to 90-120 km or to 90-200 km, both of which intersect with the "transition region" identified here. In keeping with this situation, this paper will give reasonable, and not precise, values for the bounding altitudes.

### 1.3. Physical Basis, Thermospheric Concepts, and Models

The physical basis of the static composition model lies in the treatment of the thermosphere as a coupled multicomponent fluid, which in turn follows directly from kinetic theory (e.g., Schunk and Nagy [2000], Chapman and Cowling [1970], abbreviated as CC70). In the presence of dissipation and in the absence of external forcing, treated either as internal sources or as disturbances at the boundary, the static approximation should ideally represent an average or stationary solution on short time scales (e.g., hourly) [Picone et al., 2013], and should represent an asymptotic solution for capturing variations of the composition and temperature on longer time scales (e.g., seasonal) via harmonic expansion of key parameters,

The discussion will include two candidate representations of a static fluid: (1) separate hydrostatic balance of each species and (2) so-called diffusive "equilibrium" of the species. While the latter candidate is standard to thermospheric terminology, the precise definition includes nonzero thermal diffusion, which precludes a consistent static model. As a result, the term "diffusive equilibrium (DE)" is incorrect and misleading with regard to defining a static core representation of empirical thermospheric models.

For example, the National Center for Atmospheric Research (NCAR) family of Thermospheric General Circulation Models (most recently the TIME−GCM [Roble and Ridley, 1994], where, I, E, and M denote "Ionosphere," "Electrodynamics," and " Mesosphere") uses species by species hydrostatic balance to define an upper boundary condition for major neutrals but uses the descriptor "diffusive equilibrium" (e.g., Dickinson et al. [1984]). While DE indeed reduces to species hydrostatic balance when thermal diffusion is zero (see below), this unfortunately supports the impression [Walker, 1965] that general DE condition is in fact the correct condition to define a static model. As a result, the Mass Spectrometer Incoherent Scatter (MSIS$^®$-) class of empirical models has used DE with nonzero thermal diffusion to define the density profiles of minor species. Other empirical thermospheric models also use DE with nonzero thermal diffusion [Berger et al., 1998; Bruinsma, 2015; Jacchia, 1971; Bowman, 2008].The next MSIS$^®$-class model will remove thermal diffusion from representations of all species density profiles.

Our motivation, therefore, is not only a fundamental theoretical interest in capturing the essence of thermospheric physics but also applications, including development of next generation empirical models, assimilation or inversion of data, and evaluation and initialization of complex physics-based simulation codes. The thermospheric database will continue to be sparse when distributed across time and space. Any assimilation code or empirical model must therefore use physics to fill these data gaps and to represent temperature, species density, and mass density data self-consistently. Static thermospheric models have thus proved useful and perhaps essential in constructing forward models for thermospheric data analysis [Meier et al., 2015].

A static model is clearly a very low order approximation to the complex dynamic thermosphere that is observed. Nevertheless, this representation is physically meaningful for seasonal or longer time scales [Picone et al., 2013] and for developing a physical picture of a thermospheric state. A closely related analog is the use of hydrostatic balance as the lowest−order approximation to the continuity equation governing conservation of fluid momentum (e.g., Peixoto and Oort [1992]).

Rather than appealing to hydrostatic balance equations as low−order approximations to the rigorous conservation equations for species momenta (or equivalently Newton's Second Law), we instead use the rigorous equations to investigate dynamically static models, in which the species velocities and their time derivatives are identically zero. Species hydrostatic balance then follows naturally from the species equations of motion. This approach is necessary in order to deal with the intrusion of diffusive equilibrium into the formulations of past static empirical models of the thermosphere. While diffusive equilibrium, properly defined, might have applications to thermospheric physics, this condition has no place in a physically faithful formulation of a static species altitude profile.

### 1.4. Paradigm for Current Empirical Models

A paradigm for static thermospheric models is the "Bates−Walker" profile [Walker, 1965] representation of the MSIS® (Mass Spectrometer Incoherent Radar)-class model, of which the latest version is the Naval Research Laboratory Extended MSIS® model (2000), or "NRLMSISE-00" [Picone et al., 2002]. This model seeks a physically realistic baseline composition profile by assuming so−called "diffusive equilibrium" near the exobase, a so-called "fully-mixed" composite fluid [Chamberlain and Hunten, 1987, pp. 4, 90] near the mesopause, and a smooth interpolation over altitude in the transition between the two regions. The transition, while smooth, is a nonphysical and nonlinear interpolation between the two limiting physical representations. The formulation then constrains the total mass density to approximate overall hydrostatic balance by including artificial data in the fitting process. For each species, altitude-dependent parametric multiplicative factors modify the above baseline profile to account for chemistry, dynamics, and complex processes that can induce regional maxima. Aside from these ad hoc factors, the static composition profile assimilates data according to the core regional physics models described above.

The purely mathematical (i.e., nonphysical) interpolation of species density profiles in the transition region and the use of simple ad hoc multiplicative factors have been necessary because the extant database has been sparse and the physics incompletely characterized in the middle and lower thermosphere. Future versions of the model will capitalize on sizable new data sets that are now available for the middle and lower thermosphere from various space missions: NASA TIMED/GUVI (Thermosphere, Ionosphere, Mesosphere Energetics and Dynamics/Global Ultraviolet Imager) and /SABER (/Sounding of the Atmosphere using Broadband Emission Radiometry), (2) the Canadian Space Agency OSIRIS sensor, aboard the European Space Agency Odin Satellite (website, odin-osiris.usask.ca), and Envisat/MIPAS, the European Space Agency's (ESA) Environmental Satellite/ Michelson Interferometer for Passive Atmospheric Sounding.

This paper presents the result of our inspection of the core MSIS™ formulation and of baseline static models of thermospheric composition and temperature profiles. Physical drivers (solar, geomagnetic) and spatial and temporal variations naturally occur in NRLMSISE−00 as perturbations (regression, harmonic, spherical harmonic) [Hedin, 1987; Picone et al., 2013] of the parameters of the static profile. To generate the MSIS®− class models, one uses a maximum likelihood procedure to evaluate the many parameters optimally from the available thermospheric database. Explicit time scales range from daily to seasonal, and implicit time scales derive from proxies for the solar and geomagnetic drivers.

### 1.5. Outline

The desired baseline static composition profiles follow from the governing equations of fluid dynamics. The discussion below therefore begins with the key differential equations describing Newton's Second Law and conservation of mass. (As an aside, from them follows conservation of momentum). In seeking a more physically faithful model of thermospheric composition, the analysis has revealed some problems with terminology and concepts that bear a similarity to urban legend in the community. Terms like "fully-mixed fluid" and "diffusive equilibrium" are of secondary relevance or are inappropriate for defining static models as limiting cases of the differential equations governing thermospheric dynamics. Instead, the strength of interspecies coupling is the key factor that renders the thermosphere into vertical regions.

Section 2 defines terms and presents the multifluid equations of motion, in which interspecies coupling plays a prominent role. Section 3 applies the dynamical equations to regions of "weak" (negligible) and "strong" (dominant) coupling among species momenta and to represent the intervening region of transition between the two extremes. Section 4 defines terminology related to static approximations and then derives static models of the three vertical regions, based on the applicable dynamical equations from Section 3. Section 5 discusses the concept of diffusive equilibrium and shows that such a constraint is inconsistent with a static species density profile (Section I.2). Section 6 summarizes results. Appendix A provides an example of a closed−form static solution for the equations of Section 4 in the middle and upper thermosphere (including a transition region) where the Bates temperature profile is a physically consistent representation. The results in Appendix A will find application in the next−generation MSIS® model (viz., Picone et al. [2013]). Appendix B briefly explores the relationship between interspecies momentum transfer and diffusion.

## 2. Fundamental Fluid Equations and Interspecies Coupling

As discussed earlier, the desired static composition profile follows from the basic dynamics of an ideal fluid, i.e., one which is non-viscous, and further excludes heat sources or sinks. One then considers the thermosphere as a multicomponent fluid, following a set of differential equations implementing Newton's Second Law for each of N components or species (especially Schunk and Nagy [2000]; to a lesser extent, CC70, Section 6.63; CC58, Note I):

$$\left(\frac{\partial \mathbf{v}_i}{\partial t} + \mathbf{v}_i \cdot \nabla \mathbf{v}_i\right) = -\frac{1}{\rho_i}\nabla p_i + \mathbf{g}_i + \frac{1}{\rho_i}\sum_j \omega_{ij}(\mathbf{v}_j - \mathbf{v}_i)\,;\, i, j \in \{1, 2, \ldots, N\}. \quad (2.1)$$

Unless stated otherwise, we use CGS units. In equation (2.1), for position $\mathbf{r}$ and time t, the indices i and j identify species; $\mathbf{v}_i(\mathbf{r}, t)$ is the velocity of the $i^{th}$ species; $\rho_i(\mathbf{r}, t)$ is its mass density; $p_i(\mathbf{r}, t)$ is its partial pressure, $\mathbf{g}_i(\mathbf{r}, t)$ is the acceleration due to external, noncollisional forces on species i; and $\omega_{ij}$ is an interaction frequency for momentum transfer between species i and species j. The mass density is

$$\rho_i(\mathbf{r}, t) = m_i\, n_i(\mathbf{r}, t)\,;\, i \in \{1, 2, \ldots, N\}, \quad (2.2)$$

in which $m_i$ is the particle mass and $n_i(\mathbf{r}, t)$ is the local number density of species i. The partial pressure is

$$p_i(\mathbf{r}, t) = n_i(\mathbf{r}, t)\, k_B\, T(\mathbf{r}, t)\,;\, i \in \{1, 2, \ldots, N\}, \quad (2.3)$$

where $k_B$ is Boltzmann's constant and $T(\mathbf{r}, t)$ is the local temperature. Note in passing that equation (2.3) assumes that all species are in local thermal equilibrium at temperature T. The discussion subsequently assumes that the pressure gradient is vertical and the external force is gravity, for which $\mathbf{g}$ is species independent.

The last term in equation (2.1) denotes interspecies coupling or momentum transfer, sometimes called "interspecies drag." CC58 (Note I) and CC70 (Sections 6.62−6.63) offer a candidate approximate expression for $\omega_{ij}$:

$$\omega_{ij} \approx \frac{p_i p_j}{p D_{ij}}, \qquad (2.4)$$

where $D_{ij}$ is an approximate multicomponent diffusion coefficient for species i and j. This connects interspecies coupling to molecular diffusion and thereby to the concept of diffusive equilibrium, discussed in Section 4. While the qualitative derivation of equation (2.4) by CC70 is reasonable, one may combine equation (2.1) with a diffusion equation, (5.1.1), derived rigorously by CC70, to infer a physical inconsistency with the above approximation (Appendix B).

In the absence of chemical or photochemical reactions, the mass and number of each species i are conserved according to a continuity equation

$$\left(\frac{\partial \rho_i}{\partial t} + \nabla \cdot \rho_i \mathbf{v}_i\right) = 0 \,;\, i \in \{1, 2, \ldots, N\}. \qquad (2.5)$$

Combining equations (2.1) and (2.5) gives the continuity equation for conservation of species momentum of an ideal fluid. In this way the pair of equations for conservation of species mass (or particle number) and momentum is equivalent to the combination of equations (2.1) and (2.5). Then equation (2.1) implies conservation of species momentum and equation (2.6) below implies conservation of total momentum.

Throughout the thermosphere, Newton's Second Law also holds for the composite ideal fluid,

$$\left(\frac{\partial \mathbf{v}_0}{\partial t} + \mathbf{v}_0 \cdot \nabla \mathbf{v}_0\right) = -\frac{1}{\rho}\nabla p + \mathbf{g} \,. \qquad (2.6)$$

Here the total number density, mass density, fluid velocity, and pressure as functions of location are (CC70], Section 2.6)

$$n(\mathbf{r}, t) = \Sigma_i n_i(\mathbf{r}, t)\,, \qquad (2.7)$$

$$\rho(\mathbf{r}, t) = n(\mathbf{r}, t)\, m(\mathbf{r}, t) = \Sigma_i \rho_i(\mathbf{r}, t) = \Sigma_i n_i(\mathbf{r}, t)\, m_i\,, \qquad (2.8)$$

$$\mathbf{v}_0 = [\Sigma_i \rho_i(\mathbf{r}, t)\mathbf{v}_i]/\rho(\mathbf{r}, t)\,, \qquad (2.9)$$

and

$$p(\mathbf{r}, t) = n(\mathbf{r}, t)\, k_B\, T(\mathbf{r}, t) = \Sigma_i p_i(\mathbf{r}, t) = k_B\, T(\mathbf{r}, t)\Sigma_i n_i(\mathbf{r}, t)\,, \qquad (2.10)$$

where the mass per particle of the composite fluid is $m(\mathbf{r})$, which follows from equations (2.7) and (2.8).

Conservation of mass also holds throughout the thermosphere, so that

$$\left(\frac{\partial \rho}{\partial t} + \nabla \cdot \rho \mathbf{v}_0\right) = 0\,. \qquad (2.11)$$

**3. Thermospheric Dynamical Regions**

Equation (2.1) allows the identification of altitude regions based on the ideal dynamics (as specified above) of a nonreactive, nondriven thermosphere. This corresponds to a "background" or "baseline" thermospheric state. To develop a static representation of the baseline thermospheric state, we investigate the physical properties of each altitude region within the limitations of the dynamical model, and, from this, we specify regional differential equations defining local static thermospheric composition profiles. Assuming smoothness and continuity across regional boundaries and specifying a thermospheric temperature profile then allows one to derive the desired physically consistent, approximate solutions representing static composition profiles.

The discussion assumes elements of qualitative thermospheric physics demonstrated experimentally, as inferred by NRLMSISE-00 from the extant database, and shown by physics-based model simulations (e.g., Roble et al. [1987]). In particular [Meier et al., 2001], the Bates temperature profile is appropriate in the upper thermosphere, and species density decreases exponentially as altitude approaches the exobase. Since the discussion involves local, 1D altitude profiles of the variables, the state variables in equations (2.1)-(2.10) vary only with time and altitude z.

Fortunately, a complete, closed-form, time-dependent, approximate solution of the general equations in Section 2 is not necessary to accomplish the task, since the target application of interest here is a next−generation empirical model of thermospheric climate (cf. [Picone et al., 2013]). For such a model, the parameters of our core static composition profile will themselves depend on parametric temporal and spatial variations and on parameteric regression functions of proxies for geophysical drivers. The values of these parameters will follow from the underlying database. To study the dynamical, driven, reactive thermosphere, the community employs comprehensive numerical simulation (general circulation) codes, e.g., Roble and Ridley [1994] to compute approximate solutions for specific initializations.

### 3.1. Weak (Negligible) Interspecies Coupling Region, $z \geq z_u$

Above some altitude, denoted $z_u$, the interspecies coupling, in equations (2.1) - (2.4) decreases to a negligible value. This "weak coupling limit" of the fluid, consists of noninteracting species, each satisfying equation (2.1) without the interspecies drag term i.e.,

$$\left(\frac{\partial \mathbf{v}_i}{\partial t} + \mathbf{v}_i \cdot \nabla \mathbf{v}_i \right) \approx -\left(\frac{1}{\rho_i}\frac{dp_i}{dz} + g\right)\mathbf{e}_z \, ; i \in \{1, 2, \ldots, N\}, z \geq z_u \, , \qquad (3.1.1)$$

where $\mathbf{e}_z$ is the local unit altitude vector. In the weak-coupling limit, the species velocities differ and, loosely speaking, the various species do not "see" each other. For this reason, this paper will treat the term "uncoupled" as qualitatively equivalent to "weakly coupled," at least in the sense of a limit as the coupling factor, $\omega_{ij}$, approaches zero with increasing altitude.

As an aside, although this discussion does not consider the influence of the flow on the temperature, the limit of zero coupling carries similar concerns regarding the validity of thermal equilibrium of the mixture at a single temperature T as altitude increases. Here relaxation to thermal equilibrium would occur on longer time scales with higher altitude and individual species could conceivably have different characteristic temperatures. We do not seek to address such possibilities here and are instead investigating the implications of the standard thermospheric model used in both empirical representations and detailed simulation codes.

Below (Section 3.3), the velocity $v_i^w(z, t)$ denotes the solution of equation (3.1.1), i.e., the weak coupling value at any given altitude z, even below $z_u$.

## 3.2. Strong Interspecies Coupling Region, $z \leq z_\ell$

At altitudes below a boundary denoted $z_\ell$, the interspecies collision frequency becomes large, producing a strongly coupled multifluid, so that interspecies drag dominates. That is, the solution to the set of equations (2.1) is approximately

$$\mathbf{v}_i(z, t) \approx \mathbf{v}_j(z, t) \approx \mathbf{v}_0(z, t); \; i, j \in \{1, 2, \ldots, N\}; \; z \leq z_\ell, \quad (3.2.1)$$

where $\mathbf{v}_0$ is the mass−velocity [CC70, Section 2.5] of the composite fluid, so that only Newton's Second Law for the composite ideal fluid, equation (2.6) remains independent.

Following equation (3.2.1), the constituents of a fluid particle at a given location and time travel together and have done so throughout the evolution. Because a single dynamical equation describes the fluid, one concludes that, without chemical reactions or other species-dependent drivers, the individual species are indistinguishable in this region. The fact that single nonreactive fluid conservation equations for mass, momentum, and energy apply to this region means that one may treat the effective mass per constituent particle as constant, independent of species, without loss of generality. Furthermore, in the absence of chemistry and photochemistry, equation (2.7) and the continuity of the state vector at the upper boundary, $z_\ell$, of this region imply that the average mass per particle in this region is a constant equal to the average particle mass m at the boundary:

$$m(z, t) = m_\ell = [\Sigma_i \, n_i(z_\ell, t) \, m_i] / n(z_\ell, t) = \text{constant}; \; z \leq z_\ell. \quad (3.2.2)$$

For each species, "i," substituting equations (3.2.1), (2.6), and (3.2.2) into equation (2.1) gives us

$$\left(\frac{\partial \mathbf{v}_i}{\partial t} + \mathbf{v}_i \cdot \nabla \mathbf{v}_i\right) = \left(\frac{\partial \mathbf{v}_0}{\partial t} + \mathbf{v}_0 \cdot \nabla \mathbf{v}_0\right) = -\frac{1}{\rho}\nabla p + \mathbf{g} = -\frac{1}{n_i m_\ell}\nabla p_i + \mathbf{g}\,; \; i \in \{1, 2, \ldots, N\}; \; z \leq z_\ell. \quad (3.2.3)$$

An alternative, straightforward inference from equation (3.2.3) is that, in a strongly coupled multifluid (first equality), each species flow is characterized by the same effective particle mass ($m_\ell$, third equality) if and only if the local mixing ratio is constant for $z < z_\ell$. The local mixing ratio is the ratio of species number density to total number density at altitude z, so that

$$n_i(z, t) / n(z, t) = n_i(z_\ell, t) / n(z_\ell, t) = \text{constant}\,; \; z \leq z_\ell. \quad (3.2.4)$$

Thus, in the absence of reactions, the mixing ratio is constant and each particle of each species "i" moves as though an average particle of "effective mass" $m_i^e(z) = m_\ell$, which does not vary with position in the single-fluid region. One may then substitute $m_i^e(z, t)$ for $m_\ell$ in equations (3.2.3) to provide a limiting functional representation for the governing equations within the transition region (Section 3.3 below) as z approaches $z_\ell$ from above. Within the transition region (next subsection), $m_i^e(z, t)$ will not be constant, instead transitioning smoothly from a value of $m_i$ to $m_\ell$ as the altitude decreases from $z_u$ to $z_\ell$.

The strong coupling limit is commonly called a "fully-mixed fluid" or the "mixed state;" see for example, Chamberlain and Hunten [1987], p. 4. The latter terms lack a precise mechanistic description and imply a qualitative picture that is unnecessary to the derivation of equation (3.2.1). Instead, the existence of a region that is bounded above by an altitude "$z_\ell$" and in which the mixture acts as single fluid does not depend on mixing at all. Consequently, one must conclude that the more physically faithful qualitative description of this regime is the "single fluid limit" or "single fluid region."

On the other hand, the level of mixing, via turbulence or instabilities, can be important in a quantitative sense via the treatment of eddy diffusivity as either (1) a local enhancement of molecular transport processes, including diffusion (e.g., [Dickinson et al., 1984]), or, alternatively, as (2) an effective coefficient for transport of the single fluid region upward in altitude (e.g., Chamberlain and Hunten [1987], p. 90ff, especially equations (2.3.10)−(2.3.11)). If one views the eddy diffusivity as an enhancement of molecular transport processes (item (1) in previous sentence) and applies equation (2.4) naively, substituting a larger diffusion coefficient for the molecular diffusion coefficient, $D_{ij}$, the effective interaction strength $\omega_{ij}$ would be smaller at a given altitude, so that the single fluid region (upper boundary $z_\ell$) would move lower in altitude.

If local, small scale motions are viewed in the latter sense (item (2) above), i.e., as a mechanism for enhanced transport within, and at the boundary of, the single fluid region, the eddy diffusivity can dissipate nonuniform local structure within the lower thermosphere and transport the single fluid region (i.e., upper boundary $z_\ell$) higher in the thermosphere, as explained qualitatively by Chamberlain and Hunten, and by Section 4, below in this paper. Either way, the precise value of the boundary, $z_\ell$, would then depend on the level of turbulence, as induced by various fluid instabilities (including breaking gravity wave phenomena). Nevertheless, the existence of such a (single fluid) region does not depend on the level of local, small scale motion, instead being determined by the strength of interspecies momentum transfer.

### 3.3. The Transition Region, $z_\ell < z < z_u$

We return to equation (2.1) at some fixed geographical location (latitude $\theta$, longitude $\phi$) to consider the time− and altitude−varying transition from the single fluid region to the uncoupled multifluid domain. In the absence of local drivers, this transition is necessarily smooth and is continuous at the boundaries. Here, the interspecies drag (momentum transfer) term is neither dominant nor negligible, so that species i moves with a momentum/per particle of $m_i v_i$ that is intermediate between the effective momentum, $m(z, t) v_0(z, t)$, of the species in the composite fluid under strong coupling and $m_i v_i^w(z, t)$, the weak coupling solution corresponding to equation (3.1.1) and independent of other species momenta.

Given the lack of external drivers or perturbing boundary conditions, this subsection develops an approximate background solution that implements the picture in CC70 Sections 6.62−3, in which interspecies momentum transfer causes the velocity of each species i to relax over time to the composite fluid velocity $\mathbf{v}_0(z, t)$ with governing time scale $\tau_i(z, t)$. This is in fact the same type of single fluid state that occurs in the strong coupling region. In this simple picture, with increasing altitude in the transition region, $\tau_i(z, t)$ increases toward infinity, since the interspecies coupling decreases toward zero.

The realization that, in the single fluid (strong coupling) region, each species acts as though it has an effective mass $m_\ell$ and moves at the same (composite) velocity, $\mathbf{v}_0$, gives us a method of interpreting the solution in the transition region and representing the solution approximately. That is, one may account for interspecies coupling in the transition region by representing the solution as interpolating between the two limits, so that the species move as though intermediate between single fluid and noninteracting multifluid properties. This interpolation, for each species i, leads us to define approximate profiles of effective mass per particle $m_i^e(z)$ and effective velocity per particle, $\mathbf{v}_i^e(z)$ that are continuous at the transition region boundaries.

### 3.3.1. Transition Region Model

Again since the primary interest is local altitude profiles, the following suppresses geographic coordinates in the functions. A convenient representation (see below) is a simple parametric altitude interpolation of the velocity,

$$\mathbf{v}_i(z, t) \approx \mathbf{v}_i^e(z, t) \approx \mathbf{v}_i^W(z, t) + \beta_i(z, t)[\mathbf{v}_0(z, t) - \mathbf{v}_i^W(z, t)] \; ; \; z_\ell < z < z_u, \qquad (3.3.1)$$

and a related interpolation of the effective mass per particle, consistent with the form of equation (3.2.3), which has the effective particle mass in the denominator, i.e.,

$$1/m_i^e(z, t) \approx 1/m_i + \gamma_i(z, t)[1/m(z, t) - 1/m_i]\} \; . \qquad (3.3.2)$$

In equation (3.3.1), $\mathbf{v}_i^W$ is a solution to the weak coupling equation, (3.1.1), at altitude z. As indicated below (Section 4.3) other interpolation schemes for $m_i^e$ are reasonable and might be more convenient. Here the interpolation coefficients $\beta_i(z, t)$ and $\gamma_i(z, t)$ are perfectly general and together represent solutions to the governing differential equations [see equation (3.3.2.1) and second paragraph, next subsection] for given initial and boundary conditions. For any time, t, the coefficients satisfy the usual interpolation constraints:

$$z \to z_u \Rightarrow \beta_i, \gamma_i \to 0 \; ; \; z \to z_\ell \Rightarrow \beta_i, \gamma_i \to 1. \qquad (3.3.3)$$

However, in this model the coefficients also account for relaxation induced by the coupling term in equation (2.1). Following the suggestion by CC70, Sections 6.62−3 one may, for example, define a relaxation time as

$$\tau_i(z) = \begin{cases} 1/\max\{\omega_{ij}(z) \,|\, j \text{ ranging}\} & \max\{|\mathbf{v}_j(z) - \mathbf{v}_i(z)| \; ; \; j \text{ ranging}\} > 0 \\ 0 & \mathbf{v}_j(z) \approx \mathbf{v}_0(z) > 0 \; ; \; j \text{ ranging} \end{cases}, \qquad (3.3.4)$$

so that for a nonzero velocity field at time $t_0$, $\tau_i(z, t_0) \to 0$ for $z \to z_\ell$ and $\tau_i(z, t_0) \to \infty$ for $z \to z_u$. For a static field, with velocity and acceleration zero everywhere, the coupling term is zero and the relaxation time may be considered infinite, since the resulting static solution is unchanged over time. This is relevant to the particle mass interpolation but not the velocity interpolation (see last sentence of this subsection).

For a nonzero relaxation time, one may define

$$\beta_i(z, t) = \beta_i(z, t_0) \, B_{\beta i}([t - t_0]/\tau_i(z, t_0)) \; , \qquad (3.3.5)$$

and

$$\gamma_i(z, t) = \gamma_i(z, t_0) \, B_{\gamma i}([t - t_0]/\tau_i(z, t_0)) \; , \qquad (3.3.6)$$

where the functions $B_{\beta i}$ and $B_{\gamma i}$ ensure that the coefficients $\beta_i$ and $\gamma_i$ relax over time to the single fluid representation in the interior of the transition region, i.e., for finite relaxation time $\tau_i(z, t_0)$,

$$t \to \infty \Rightarrow B_{\beta i} \to 1/\beta_i(z, t_0) \text{ and } B_{\gamma i} \to 1/\gamma_i(z, t_0), \; z_\ell < z < z_u \,, \qquad (3.3.7)$$

For a very long relaxation time ($\tau_i(z, t_0) \to \infty$), $B_{\beta i}([t - t_0]/\tau_i(z, t_0)) \to 1$ and $B_{\gamma i}([t - t_0]/\tau_i(z, t_0)) \to 1$, so that $\beta_i(z, t) \to \beta_i(z, t_0)$ and $\gamma_i(z, t) \to \gamma_i(z, t_0)$, applicable to a static field or, alternatively, to a dynamical field near the upper boundary of the transition region.

As a simple example, consider the temporal interpolations

$$B_{\beta i} \approx [1 - \exp\{-[t - t_0]/\tau_i(z, t_0)\}]/\beta_i(z, t_0) + \exp\{-[t - t_0]/\tau_i(z, t_0)\} \qquad (3.3.8)$$

and
$$B_{\gamma i} \approx [1 - \exp\{-[t - t_0]/\tau_i(z, t_0)\}]/\gamma_i(z, t_0) + \exp\{-[t - t_0]/\tau_i(z, t_0)\}. \quad (3.3.9)$$

At the upper boundary of the transition region, where the coupling is very weak, equations (3.3.7)−(3.3.9) show that, for finite t, the corresponding very long relaxation time ($\tau_i(z, t_0) \to \infty$) gives us $\beta_i(z, t) \to 0 = \beta_i(z_u, t_0)$ and $\gamma_i(z, t) \to 0 = \gamma_i(z_u, t_0)$, consistent with equation (3.3.3). Similarly, near the lower boundary of the transition region, with a relaxation time ∼ 0, $\beta_i(z, t) \to 1 = \beta_i(z_\ell, t_0)$ and $\gamma_i(z, t) \to 1 = \gamma_i(z_\ell, t_0)$, again consistent with equation (3.3.3). In the interior, where the relaxation time is finite, the coefficients relax toward unity (the value for a single fluid state) over time, increasingly slowly as altitude increases or more quickly over time as altitude decreases. This is the effect envisioned by CC70. For a static solution, with an effectively infinite relaxation time, $B_{\gamma i} = 1$, so that the mass interpolation coefficient is constant in time, i.e., $\gamma_i(z, t) = \gamma_i(z, t_0)$; the velocity interpolation term is not meaningful since the velocity field is zero.

### 3.3.2. Interpretation of Unperturbed Transition Region Representation

Our purpose in Section 3 is to lay the physical groundwork and to provide a closed−form representation of the temporal evolution of an unperturbed thermospheric state, from which follows the static model presented in Section 4. To do this, we have worked from the time dependent equations of motion (2.1) and (2.6). In our transition region model, equations (3.3.1) and (3.3.2) for the approximate velocity and effective particle mass, and the relaxation conditions (3.3.3)−(3.3.7) account for the effect of the interspecies coupling term in equation (2.1) over the time interval [$t_0$, t], where the arbitrary initial time is $t_0$ and the current time is t. Ideally, the species velocity satisfies dynamical equation (2.1):

$$\left(\frac{\partial \mathbf{v}_i^e}{\partial t} + \mathbf{v}_i^e \cdot \nabla \mathbf{v}_i^e\right) = -\frac{1}{n_i m_i^e}\nabla p_i + \mathbf{g}_i + \frac{1}{n_i m_i^e}\sum_j \omega_{ij}(\mathbf{v}_j^e - \mathbf{v}_i^e); \ i, j \in \{1, 2, \ldots, N\}, \quad (3.3.2.1)$$

in which all quantities, except **g** vary with both time and location. While this equation is not practical for numerical simulations, the equation does provide the basis for the heuristic discussion (below) of our transition region representation of the time dependent background (unperturbed) thermospheric fluid state. From this should follow the physics underlying the static models that we seek. Ultimately, for the static solution, the LHS and the coupling term are zero (since the velocity and acceleration fields are zero); the interpolation coefficients are constant in time; and the pressure gradient and gravity terms on the RHS determine the species number density profiles (Section 4).

In a time−dependent thermospheric state, the interpolation coefficients, $\beta_i(z, t)$ and $\gamma_i(z, t)$, defining $\mathbf{v}_i^e(z, t)$ and $m_i^e(z, t)$ (noting that geographic coordinates are suppressed) should embody the transition region solution, given (a) the temperature profile $T(\mathbf{r}, t)$ and the species number density profiles, $\{n_i(\mathbf{r}, t)\}$, to construct the right hand side (RHS) of equation (3.3.2.1), and (b) equations (2.6) and (3.1.1), which provide the time derivatives of the velocity fields (for strong and weak coupling) for the left hand side (LHS) of equation (3.3.2.1). Then equation (3.3.2.1) becomes a theoretical equation relating the interpolation coefficients.

In this model, for an altitude profile varying with z and t, the effective velocity $\mathbf{v}_i^e(\beta_i, z, t)$ and the effective mass $m_i^e(\gamma_i, z, t)$ specify the dynamical state at time t. In particular, the coefficients $\beta_i(z, t)$ and $\gamma_i(z, t)$ account for the relaxation in altitude and time toward a single fluid state over the interval [$t_0$, t] within the interior of the transition region. At early times or on short time scales $[t - t_0]/\tau_i(z, t_0) \sim 0$, the coefficients contain the effects of decoupling, with increasing altitude, in the transition of the background state from the single fluid state of the lower thermosphere to the weakly coupled state of the upper thermosphere.

To demonstrate the details of coupling and relaxation, consider time t and set $\beta_i(z, t) \equiv \beta(z, t) \equiv \beta$ independent of species. Performing the substitutions, the coupling term of (3.3.2.1) for each species "i" causes relaxation to the single fluid state for time t' > t according to

$$\left[\frac{d^e \mathbf{v}_i^e}{dt}\right]_{rel} \sim \frac{1-\beta(z,t)}{n_i m_i^e} \sum_j \omega_{ij}(\mathbf{v}_j^w - \mathbf{v}_i^w) \; ; \; i, j \in \{1, 2, \ldots, N\}, \qquad (3.3.2.2)$$

in which the term in brackets on the LHS is the total time derivative, corresponding to the LHS of equation (3.3.2.1), and the subscript "rel" indicates the portion of the future relaxation toward a single fluid state from time t to time t' > t. The RHS shows the altitude and time dependence of the interpolation coefficient explicitly for emphasis of its role in representing the relaxation process from time $t_0$ to time t. The relaxation after a given time t, then, depends on the portion of the velocity field that corresponds to very weak coupling, i.e., a solution, at transition region altitude z and time t, to equation (3.1.1), which also governs the dynamics of the upper thermosphere. As t gets very large for $z_\ell < z < z_u$, $\beta(z, t)$ approaches 1 and the relaxation to a single fluid state approaches completion.

To reiterate: This section discusses the evolution of the unperturbed (background) thermospheric fluid state over time, as a backdrop to the equations defining a static state or model. The above discussion, therefore, does not address the extent of the relaxation process in a realistic thermosphere (next subsection).

### 3.3.3. Realistic Situations and Applications of the Transition Region Model

A more realistic thermosphere experiences a variety of sizable perturbations, e.g., time−varying boundary conditions, solar and geomagnetic heating, photochemistry, and viscous drag. While viscous drag tends to eliminate gradients or nonuniformities in momentum, a relaxation process itself, the various perturbations and energetic sources interrupt relaxation of the background state to the single fluid state in the transition region. Under these influences, relaxation of the transition region to a static state, intermediate between (and interpolating between) the strong and weak coupling regions, does not appear likely. Rather the static state arises from the equation (3.3.2.1) for the unperturbed evolution from an initial background state with velocity and acceleration fields of zero.

On the other hand, on shorter time scales than those of the perturbations (or averaging the solution over a range of shorter time scales [Picone et al., 2013]), the static state, as defined in Section 4, can be a meaningful approximation to a snapshot of the solution of equation (3.3.2.1). This arises from the approximate hydrostatic balance observed in the atmosphere [Dutton, 1986, Section 4.1] and is a basic assumption of most current GCMs. To represent the effects of coupling in the transition region for a particular snapshot of the thermosphere, one may set $t = t_0$ in the definitions of the interpolating coefficients in equations (3.3.5)−(3.3.6). To represent climatology (longer time scales), one can parameterize $\beta_i(z, t)$ and $\gamma_i(z, t)$, allowing the internal parameters to be functions of time and heating sources to capture the various perturbations listed above. In the target application, empirical models, these internal parameters are optimal values computed from the composition (and temperature) database (next section).

As an aside, the prescription of equations (3.3.1), (3.3.2), and (3.3.2.1), along with the paradigm of Picone et al. [2013], could also provide a framework for designing a background coupled wind, composition, and temperature model. At present, this constitutes merely a heuristic picture to be implemented via considerable mathematical work and insight on the part of the model developer. Nevertheless, this approach does offer a physically faithful, mathematical framework for such an algebraic model of the thermosphere.

## 4. Static Model Approximations

### 4.1. Terminology

As stated earlier, the term "static model" implies that the velocity and acceleration fields are initially zero and remain so over time in the absence of drivers and varying boundary conditions. In this sense, the solution is "stationary." As indicated below, this leads directly to hydrostatic balance for the composite fluid and to "species hydrostatic balance." The latter condition occurs when each species, taken separately, is in hydrostatic balance with regard to the gradient of the species partial pressure. Such approximations are in frequent use because the pressure gradient and gravitational forces are often the dominant terms or cancel approximately in the respective momentum equations (see citations below).

Because of the prevalent use of so−called "diffusive equilibrium" conditions to define static thermospheric states, our discussion will necessarily include the lifetime of a given unperturbed static model or state. This does not concern fluid dynamic stability, which involves evolution after a small perturbation from a background or equilibrium state. Further, the concept of equilibria necessarily incurs considerations of stability. Perturbing a static thermosphere will necessarily render the thermosphere no longer static. For these reasons, we are not directly interested in equilibria, perturbations of a given state, or stability. To extract variations of interest from a database and to fill gaps in the database, empirical or climatological models require a physically faithful, static baseline, i.e., one which is long−lived in the absence of external sources or boundary perturbations.

Unavoidably, Section 5 considers "diffusive equilibrium," which has been the terminology and baseline of choice for empirical thermospheric models to date. Unfortunately, in such an initial state, as currently defined, the acceleration field is trivially nonzero. Therefore a thermosphere initially in "diffusive equilibrium" (velocity fields are zero) cannot remain static and is therefore not a true equilibrium, as discussed in Section I.2. For this reason, static empirical models such as the core profile representation of future MSIS®-class models, would be more transparent and physically faithful by using a formulation that excludes thermal diffusion. Empirical models to date, including the MSIS®-class models, follow the formulation of Walker [1965], adding the extra complication of nonzero thermal diffusion factors at least for minor species. This incurs, at best, an unnecessary influence on the empirical model coefficients and a less straightforward representation of the physics of various thermospheric regions, as well as similar confusion caused by the term "equilibrium."

### 4.2 Weak Coupling and Strong Coupling Regions

This discussion assumes a one-dimensional model; i.e., the fluid properties vary only with altitude. First consider the limits of the fluid equations as the altitude approaches the thermospheric boundaries, that is, in the weak and strong coupling regions. Near the exobase, the species momenta become uncoupled ($\omega_{ij} \approx 0$), so that each equation (2.1) apparently describes a separate fluid, although a constraint remains, in the form of the composite fluid equation of motion, (2.6). For an initially static thermosphere, i.e., one of in which the velocities and accelerations are initially zero, equations (2.1) reduce to initial "species hydrostatic balance,"

$$\frac{1}{\rho_i}\frac{d}{dz}p_i = -g\,;\;\;i \in \{1, 2, \ldots, N\};\,z \geq z_u, \qquad (4.2.1)$$

in the vertical direction. Then for later times, according to equations (2.1), the accelerations and velocities remain zero, so that equations (4.2.1) continue to hold, describing a state that is truly static. The same set of equations also defines a state of hydrostatic force balance, the lowest order approximation to equation (2.6) that is observed to hold in the atmosphere (e.g., Dutton [1986], Section 4.1; Peixoto and Oort [1992], Chapter 3).

Near the mesopause, the species momenta become strongly coupled, following equation (3.2.1), so that the mixture acts as a single fluid, following equations (2.6) and (3.2.3). Again, when the initial species velocities and accelerations in the strong coupling region are all zero and the dominant pressure gradient is vertical, the equations of motion (3.2.3) reduce initially to

$$\frac{1}{n_i m_\ell}\frac{d}{dz}p_i = -g\,;\;\;i \in \{1, 2, \ldots, N\};\,,z \leq z_\ell. \qquad (4.2.2)$$

Equations (2.6) and (3.2.3) then ensure that, over time, the regional solution to equation (4.2.2) continues to hold, remaining static, physically faithful, and, again, stationary. In the lower thermosphere, equation (4.2.2) also describes hydrostatic balance, the lowest order approximation to the governing equation (2.6) that applies to the atmosphere in general. As indicated in Section 3, equation (4.2.2) is equivalent to the statement that the mixing ratio of each species is constant with altitude in this region.

**4.3. Transition Region**

In the transition region, the dynamical coupling term among species is not negligible, so that the composite flow has characteristics intermediate between the extremes of uncoupled (or weakly coupled) multispecies flows and a (strongly coupled) single fluid state. In this sense, the condition of continuity at the upper and lower boundaries allows an "effective" model or picture of the transition region flow in terms of the two contributing components. The heuristic treatment in Section 3 provides an approximate separation of these two components.

The static model follows from setting the initial velocity and acceleration terms in equation (3.3.2.1) to zero:

$$\frac{1}{n_i m_i^e(z)}\frac{d}{dz}p_i = -g\,;\;\;i \in \{1, 2, \ldots, N\}. \qquad (4.3.1)$$

According to equation (3.3.2.1), the initial condition of hydrostatic balance, equation (4.3.1), ensures that the velocity and acceleration fields remain identically zero, producing a (stationary) static model and solution.

Generalizing equation (3.3.2) with no consequence, the interpolated mass per particle may also be

$$m_i^e(z) = m_i + \gamma_i(z)[m(z) - m_i]\};\;\;i \in \{1, 2, \ldots, N\}. \qquad (4.3.2)$$

This interpolation is faithful to the qualitative physics of the thermosphere and allows the construction of a closed−form thermospheric empirical model profile. With little loss of generality, the empirically determined, parameterized weighting function, $\gamma_i(z)$ may effectively subsume the altitude dependence of m(z), the parameterization of the mass profile, so that an alternative parameterization is

$$m_i^e(z) = m(z_\ell) + \gamma^0_i(z) [ m_i - m(z_\ell)] \},\qquad(4.3.3)$$

in which $z_\ell$ is the upper boundary of the strong coupling region, where the mean mass per particle is constant. Notice that the equation also reverses the direction of the interpolating factor $\gamma^0_i(z)$, so that its value is unity at $z = z_u$ and zero at $z = z_\ell$. This is convenient for the direction of integration of equation (4.3.1) in the example presented in Appendix A.

This freedom with the interpolating factor $\gamma^0_i(z)$ allows, for example, a piecewise linear or polynomial representation of the effective species mass profile within the transition region. For some representations of the temperature profile, a judicious choice of $\gamma^0_i(z)$ will enable direct closed form or algebraic (often called "analytic") integration of (4.3.1), a particularly nice situation for constructing an empirical model species profile. Appendix A shows a linear example with a path toward a general polynomial interpolation.

Two aspects of this formulation are worthy of note. First, the above discussion appropriately avoids attributing these force balance conditions as "equilibria" instead of the particular solutions to momentum conservation that these conditions represent. Secondly, these equations correspond to physically faithful, static models or solutions appropriate for the baseline representation of empirical climatological models.

## 5. "Diffusive Equilibrium"
### 5.1. Background and General Formulation

The paper by Walker [1965] extended the formulation of the Bates temperature and density profile to include thermal diffusion by adopting a further implicit assumption that so-called "diffusive equilibrium (DE)" with nonzero thermal diffusion properly represented a static species density altitude profile. A comparison with the results of Section 4 reveals these assumptions to conflict with the inference of stationary, static representations from the governing equations of fluid dynamics. Below we shall see that the inclusion of thermal diffusion renders the state of DE to be nonstationary and therefore not a static limit to the governing equations of fluid dynamics.

While tempting to begin with the paradigm of a binary mixture, the general formulation of CC70 for gas mixtures is just as transparent for our purposes. For some points, CC70's treatment of binary mixtures (Sections 8.3-8.4), is more explicit, and will be cited below.

In the thermosphere, equation (18.3,13) of CC70 shows that

$$\mathbf{d}_i + k_{Ti} \nabla \ln T \approx \sum_j \frac{n_j n_i}{n^2 D_{ij}} (\bar{\mathbf{u}}_j - \bar{\mathbf{u}}_i) = \sum_j \frac{n_j n_i}{n^2 D_{ij}} (\mathbf{v}_j - \mathbf{v}_i),\qquad(5.1.1)$$

in which $\bar{\mathbf{u}}_i \equiv \mathbf{v}_i - \mathbf{v}_0$ is the Boltzmann−averaged peculiar velocity of CC70 (p. 44), $k_{Ti}$ is the thermal diffusion ratio of species i, e.g., CC70, equation (8.4,6), (18.3,10), and

$$\begin{aligned}d_i &= \frac{1}{p}\left\{\frac{d}{dz}p_i + \rho_i g - \frac{\rho_i}{\rho}\left(\frac{d}{dz}p + \sum_j \rho_j g\right)\right\} \\ &= \frac{d}{dz}\left(\frac{n_i}{n}\right) + \left(\frac{n_i}{n} - \frac{\rho_i}{\rho}\right)\frac{d}{dz}\log p\end{aligned},\qquad(5.1.2)$$

The equation considers variations of all quantities (including g) in altitude only and suppresses the notation "(z)". The first line includes gravitational terms that actually cancel, but which have been retained to account for the acceleration of the composite fluid [equation (2.6)]. The third term (first line, parenthetical expression) is proportional to that acceleration.

Notice that the use by CC70 of relative or peculiar velocities to derive equation (5.1.1) has limited information content regarding the species velocities $\{\mathbf{v}_i\}$, even though no such limitation is implied explicitly. That is, $\mathbf{v}_0$ is not specified and cancels in the velocity differences, so that the equation holds for species velocities or relative velocities, e.g., $\{\bar{\mathbf{u}}_i\}$. This can lead to misinterpretation of solutions for $\{\mathbf{v}_i\}$ from equation (5.1.1), given coefficients of diffusion and thermal diffusion. Stated in a different way: one might consider using another equation like CC70, equation (18.3,9), to compute individual average peculiar velocity or species velocity. Unfortunately, these equations require precise knowledge of the transport coefficients to be consistent with the equations of motion given in Section 2.

In fact, the relationship, equation (5.1.1), is an approximation based on direct averaging of the particle peculiar velocity with the species Boltzmann function, corrected to first−order. With standard treatments of diffusion and thermal diffusion coefficients in theoretical and numerical models, these velocities are not solutions to the species equation of motion (2.1). This leads to the ambiguity in the interspecies drag coefficient, as discussed in Appendix B.

In an attempt to define a static approximation ("diffusive equilibrium"), one may then set the species velocities to zero to obtain

$$d_i + k_{Ti} \frac{d}{dz} \ln T = 0 \,; \, i = 1, 2, \ldots, N \qquad (5.1.3)$$

However, equation (5.1.3) does not ensure a static state. One may obtain the same equation by setting all of the individual species velocities to the composite velocity $\mathbf{v}_0$. In this latter case, the individual species velocities are nonzero and the solution is not static. Instead, the species velocities, relative to each other, are zero, thereby following equation (3.2.1), which defines a single fluid dynamical state. Conversely, according to CC70 (Section 18.3), equation (5.1.3) implies that the relative species velocities are zero. This is relevant to equation (5.2.3) below.

Again, for variations in altitude only, equation (5.1.3) expresses the condition of so−called "diffusive equilibrium," which is applied to the thermosphere by Walker [1965], Nicolet [1968], and Chamberlain and Hunten [1987], and is subsequently used in the formulations of various MSIS®-class models and other empirical thermosphere models. Deriving those formulations requires further approximations. Finally, if $k_{Ti} = 0$ for every "i," then the equation for DE reduces to species hydrostatic balance, the correct condition defining a static solution.

## 5.2. "Standard" Approximations

The general equation for $d_i$ in equation (5.1.2) is not often used in thermospheric modeling, with some noteworthy exceptions, e.g., Fuller-Rowell et al. [1996]. Popular approximations (see below) consist of (1) setting the acceleration of the composite fluid to zero or (2) assuming that one has a multicomponent mixture which includes only a single dominant species "j" and in which the species "i ≠ j" are distinctly minor. For example, the NCAR GCM family of simulation codes apply the first assumption [Dickinson and Ridley, 1972]) to model the molecular diffusion of individual species.

Static representations in the literature (e.g., Nicolet [1968]; Chamberlain and Hunten [1987], Section 2.3) use both approximations, under which one may substitute an expression [CC, equation (8.4,8)] for a single species thermal diffusion factor $\alpha_{Ti}$,

$$k_{Ti} = \frac{n_i n_j}{n^2}\alpha_{ij} \approx \frac{n_i}{n}\alpha_{Ti}, \tag{5.2.1}$$

to obtain the "standard" equation [Picone et al., 2013],

$$\frac{d}{dz}p_i = -\rho_i g - \alpha_{Ti} p_i \frac{d}{dz}\log T. \tag{5.2.2}$$

This differs from species hydrostatic balance by the additional thermal diffusion term and is equivalent to the expression following from Chamberlain and Hunten [1987] (equation (2.3.2) and the last paragraph of Section 2.3.1):

$$\frac{1}{n_i}\frac{d}{dz}n_i + \frac{m_i g}{k_B T} + \frac{1}{T}\frac{d}{dz}T + \frac{\alpha_{Ti}}{T}\frac{d}{dz}T = 0. \tag{5.2.3}$$

In equation (5.2.3), the first three terms thus represent species hydrostatic force balance.

Notice that the condition of "diffusive equilibrium" generally precludes species hydrostatic balance, since thermal diffusion is nonzero in the thermosphere for any species in the mixture. Substituting equation (5.1.3) or the approximate variants, e.g., equation (5.2.3) into the equation of motion, (2.1), of each species "i," one obtains

$$\left(\frac{\partial \mathbf{v}_i}{\partial t} + \mathbf{v}_i \cdot \nabla \mathbf{v}_i\right) \cong \alpha_{Ti}\frac{k_B}{m_i}\frac{dT}{dz} + \sum_j \omega_{ij}(\mathbf{v}_j - \mathbf{v}_i); i, j \in \{1, 2, ..., N\}. \tag{5.2.4}$$

Setting the initial velocity field to zero shows that the acceleration of species "i" will not be zero in regions of nonzero thermal diffusion (i.e., nonzero temperature gradient). Even without any perturbation, yet with an initial species velocity field of zero, DE is a transient state and hence does not represent a stationary (static) solution. One must conclude that in the context of thermospheric composition profiles, DE does not define a physically realizable static state on the longer time scales covered by empirical models.

One must therefore conclude that the general condition of "diffusive equilibrium" is not a static solution for the thermosphere. By definitions of equilibrium cited in the introduction (Section I.2, paragraph 3), "diffusive equilibrium" is not an equilibrium state at all! Species−by−species hydrostatic balance is in fact the only physical static solution with which to interpolate data gaps and, therefore, on which to base future empirical models like those of the MSIS® class. Further, the lowest order approximate equations of motion for a multicomponent fluid consist of species-by-species hydrostatic balance as given in Section 4.

## 6. Summary

A physically faithful representation of the core thermospheric composition profile is essential for an empirical model to fill gaps in the extant database and to extract from the data thermospheric variability on daily and longer time scales [Picone et al., 2013]. Working from the coupled, ideal fluid equations of motion for the separate species in the thermosphere, one may define unambiguous static solutions based on species-by-species hydrostatic balance throughout the thermosphere. In the absence of chemical (and photochemical) reactions, the thermosphere consists of three regions defined by the strength of interspecies coupling:

(1) The lower thermosphere (herein defined as $z < \sim 100$ km) is the region of strong interspecies coupling in which the species move at the velocity of the composite fluid with an effective species particle mass equal to the average particle mass in the region. That is, one may model the lower thermosphere as a single fluid. In this situation, the mixing ratio of each species is constant in the absence of chemical reactions.

(2) The upper thermosphere (herein defined as $z > \sim 200$ km) is the region of "weak," or negligible interspecies coupling, in which the species assume unequal velocities and follow equations of motion driven primarily by the gradient in partial pressure and gravity. Here, the effective species particle mass is equal to the actual species particle mass.

(3) The transition region between the upper and lower thermosphere consists of coupled species motion in which the interspecies drag term is neither dominant nor negligible, so that the species move at different velocities from each other and from the composite fluid. However, over time and at interior altitudes, $z_\ell < z < z_u$, in an unperturbed thermosphere and with some classes of simple, passive boundary conditions, the interspecies coupling term drives the species velocities toward the composite fluid velocity, $\mathbf{v}_0$. Given the constraint that fluid variables be continuous at the region boundaries, one may also treat dynamical solutions in this region as representing a transition in effective species particle mass from the actual species particle mass in the upper thermosphere to the constant average species particle mass in the lower thermosphere. Again, the authors emphasize that this picture applies to an unperturbed thermosphere, i.e., the baseline state that we seek (see last paragraph of this section).

By setting to zero the velocity and acceleration fields in the governing equations of motion for each region of the thermosphere, one obtains static models that are continuous throughout the thermosphere. These static models apply to an unperturbed baseline thermosphere and represent hydrostatic balance rather than so-called diffusive equilibrium, which is rendered nonstationary by the presence of nonzero thermal diffusion. This essentially improves, in small ways, the formulation of the MSIS®-class composition models, which until now have followed the equations of Walker [1965]. One should expect future thermospheric empirical models to use species-by-species hydrostatic balance, which is a more physically faithful representation of the static baseline thermosphere.

For realistic thermospheres, sizable perturbations (including energetic sources) on a wide range of time scales in the transition region will likely interrupt relaxation of the species velocity fields toward a particular composite velocity $\mathbf{v}_0(z)$ in the transition region. Nevertheless, on shorter time scales than those of the perturbations (or averaging the solution over a range of shorter time scales [Picone et al., 2013]), the static state, as defined in Section 4, can be a meaningful approximation to a snapshot of the solution of equation (3.3.2.1).

**Appendix A. Example: Particle Mass Interpolation Across Transition Region**

**A.1. Formulation**

This appendix provides an example of a closed-form static solution for equation (4.3.1) across the transition region ($z_\ell < z < z_u$) using the mass interpolation of equation (4.3.3) between the species particle mass $m_i$, which applies in the weakly coupled region, $z \geq z_u$, and the average species particle mass $m(z_\ell)$, which applies in the strong coupling or single fluid region of the lower thermosphere, $z \leq z_\ell$. The selected interpolation function $\gamma_i(z)$ in equation (4.3.3) is linear:

$$\gamma^0{}_i(z) = \frac{z - z_\ell}{z_u - z_\ell}, \tag{A.1.1}$$

and the particle mass values at the transition region limits are now $m_\ell \equiv m(z_\ell)$ and $m_u \equiv m_i$ in equation (4.3.3). Section A.4 shows a straightforward path toward closed−form solutions for general polynomial interpolation.

In terms of geopotential height $\zeta$ (e.g., Chamberlain and Hunten [1987], p. 69), referenced to some altitude $z = z_0$, equation (4.3.1) becomes

$$\frac{1}{p_i}\frac{dp_i}{d\zeta} = \frac{1}{n_i(\zeta)}\frac{dn_i}{d\zeta} + \frac{1}{T(\zeta)}\frac{dT}{d\zeta} = -\frac{m_\ell g_0}{k_B T(\zeta)} - \frac{(m_u - m_\ell)g_0}{(\zeta_u - \zeta_\ell)k_B}\frac{(\zeta - \zeta_\ell)}{T(\zeta)}. \tag{A.1.2}$$

In equation (A.1.2), $k_B$ is the Boltzmann constant and $g_0$ is the gravitational acceleration at altitude $\zeta_0 \equiv 0$.

For completeness and comparison to the transition region, the example includes the weak coupling region, $z \geq z_u$ ($\zeta \geq \zeta_u$), for which $m_u = m_i$, and equation (4.2.1), or equation (4.3.1) with $m_i^e(z) = m_i$, becomes

$$\frac{1}{p_i}\frac{dp_i}{d\zeta} = \frac{1}{n_i(\zeta)}\frac{dn_i}{d\zeta} + \frac{1}{T(\zeta)}\frac{dT}{d\zeta} = -\frac{m_i g_0}{k_B T(\zeta)}, \tag{A.1.3}$$

In the strong coupling region, $z \leq z_\ell$ ($\zeta \leq \zeta_\ell$), equation (4.2.2), or equation (4.3.1) with $m_i^e(z) = m_\ell$, becomes

$$\frac{1}{p_i}\frac{dp_i}{d\zeta} = \frac{1}{n_i(\zeta)}\frac{dn_i}{d\zeta} + \frac{1}{T(\zeta)}\frac{dT}{d\zeta} = -\frac{m_\ell g_0}{k_B T(\zeta)} . \tag{A.1.4}$$

## A.2. Temperature Profile

This example presumes that, for the upper thermosphere and extending downward into the lower thermosphere, the Bates temperature profile [Bates, 1959] is a reasonable approximation [Picone et al. 2013, Section 5.2] having the virtues of realistic altitude dependence in the transition region and upper thermosphere and algebraic integrability of the inverse temperature (next section). Referenced to $\zeta = \zeta_\ell \equiv \zeta(z_\ell)$, where the subscript "$\ell$" identifies the location $z_\ell$, the Bates temperature profile is

$$T(\zeta) = T_\infty\{1 - a \exp[-\sigma(\zeta - \zeta_\ell)]\} = T_\ell\{1 - a \exp[-\sigma(\zeta - \zeta_\ell)]\}/[1-a] . \tag{A.2.1}$$

In equation (A.2.1) the dimensionless coefficient is

$$a = 1 - T_\ell/T_\infty ; \tag{A.2.2}$$

$T'_\ell$ is the derivative $dT(z)/dz$, evaluated at $z_\ell$; and the inverse temperature scale height is

$$\sigma = T'_\ell/(T_\infty - T_\ell) = \Lambda_L^{-1}(1-a)/a , \tag{A.2.3}$$

in which the length scale $\Lambda_L$ is defined by the ratio of initial values, $T_\ell/T'_\ell$. Alternatively, one can use $\sigma$, the inverse scale height itself, to define a governing length scale. In general, one need not reference the Bates Temperature profile to the upper boundary $z_\ell$ of the strong coupling region, as we have done here. That is, in equations (A.2.1)−(A.2.3), one may define the Bates profile by substituting $\zeta_{ref} \equiv \zeta(z_{ref})$ for the reference geopotential height $\zeta_\ell = \zeta(z_\ell)$, where $z_{ref} \neq z_\ell$, and similarly, by substituting $T_{ref}$ for $T_\ell$ and $T'_{ref}$ for $T'_\ell$.

## A.3. Standard Species Number Density Profile for Uncoupled Species and a Single Fluid

For the regions of weakly (or un-) coupled species flows and of the strongly coupled flow (i.e., single composite fluid), equations (4.2.1) and (4.2.2) apply, respectively, and, because the mass per particle is constant in each case, the integration proceeds in the same manner. As shown in section A.1, these equations reduce to equations (A.1.3) and (A.1.4), respectively – each a modified form of equation (A.1.2), without the rightmost (interpolation) term and with the remaining right−hand term having $m_u$ (= $m_i$) or $m_\ell$ in the numerator.

Consider first the upper thermosphere ($\zeta$ above $\zeta_u$), i.e., the region of effectively uncoupled species flows, with effective species mass $m_u = m_i$, for each "i." Integration of equation (A.1.3) is straightforward for a Bates temperature profile and is readily available in the literature, e.g., Chamberlain and Hunten [1987], p. 69, and Picone et al. [2013], Section 6:

$$n_i(\zeta) = n_{iu} \left[\frac{T_u}{T(\zeta)}\right]^{1+\gamma_i} \exp[-\Gamma_i \sigma(\zeta - \zeta_u)]; \; \zeta \geq \zeta_u , \qquad (A.3.1)$$

where $n_{iu} = n_i(\zeta_u)$ and $T_u = T(\zeta_u)$, the ratio of temperature and species scale heights is

$$\Gamma_i = (1-a)/(\sigma H_{\ell i}) \equiv 1/(\sigma H_{\infty i}) , \qquad (A.3.2)$$

and the effective species density scale height is

$$H_{\ell i} = k_B T_\ell /(m_i g_0) = (1-a)k_B T_\infty/(m_i g_0) \equiv (1-a)H_{\infty i} . \qquad (A.3.3)$$

Here $T_\ell = T(\zeta_\ell)$ is the reference temperature arbitrarily chosen in Section A.2.

Similarly, consider the lower thermosphere, i.e., at altitude $\zeta \leq \zeta_\ell$, characterized by the single composite fluid equation of motion and a constant effective particle mass equal to the average mass, $m_\ell$. This example assumes that the Bates temperature profile applies at least to the vicinity of the strong coupling region just below $\zeta_\ell$.

The solution is continuous at the boundary $\zeta_\ell$, where the second term on the right−hand side (RHS) of equation (A.1.2) is zero. One may integrate equation (A.1.4) from $\zeta_\ell$ to $\zeta$, obtaining

$$n_i(\zeta) = n_{i\ell} \left[\frac{T_\ell}{T(\zeta)}\right]^{1+\gamma_\ell} \exp[-\Gamma_\ell \sigma(\zeta - \zeta_\ell)] ; \; \zeta \leq \zeta_\ell , \qquad (A.3.4)$$

where "i" again labels the species, the ratio of temperature and species scale heights is

$$\Gamma_\ell = (1-a)/(\sigma H_\ell) \equiv 1/(\sigma H_{\infty \ell}) , \qquad (A.3.5)$$

and the effective species density scale height is

$$H_\ell = k_B T_\ell /(m_\ell g_0) = (1 - a)k_B T_\infty/(m_\ell g_0) \equiv (1 - a)H_{\infty \ell} . \qquad (A.3.6)$$

As expected, the altitude variation of each species density in the lower thermosphere is identical to that of the composite fluid.

## A.4. Transition Region Density Profile Interpolation
### A.4.1. Solution for Linear Interpolation

For $\zeta_u > \zeta > \zeta_\ell$, both terms on the far RHS (second equality) of equation (A.1.2) are nonzero. Equation (A.3.4) gives the integral of the first of those terms, for the effective species particle mass $m_\ell$, combined with the integral of temperature term in the first equality.

The integral of the second term of the RHS, which interpolates between $m_\ell$ and $m_u = m_i$ for species i, and which is denoted by $n_{i2}(\zeta)$, proceeds as with the first term, but is more complex, given the additional dependence on $\zeta$:

$$\log n_{i2}(\zeta) = -\frac{g_0}{k_B} \frac{m_u - m_\ell}{\zeta_u - \zeta_\ell} \int_{\zeta_\ell}^{\zeta} \frac{\zeta' - \zeta_\ell}{T_\infty \{1 - a \exp(-\sigma[\zeta' - \zeta_\ell])\}} d\zeta'. \tag{A.4.1}$$

The approach of this section to direct integration produces a closed−form result that, while cumbersome to implement in a code, should be documented. The next subsection provides an alternative series solution that is convergent by inspection, is far easier to code, and generalizes to higher−order interpolation polynomials.

First, move the constant $T_\infty$ outside the integral and then transform variables twice: first, to the linear variable $\zeta'' \equiv \zeta' - \zeta_\ell$, whose domain is $[0, \zeta - \zeta_\ell]$, and then, after factoring $\exp(-\sigma\zeta'')$ from the denominator, to the variable

$$u \equiv \exp(\sigma\zeta'') - a \tag{A.4.2}$$

with

$$du \equiv \sigma \exp(\sigma\zeta'') \, d\zeta''. \tag{A.4.3}$$

Noting that

$$\zeta'' = \sigma^{-1} \log(u + a) \tag{A.4.4}$$

and ignoring the multipliers outside the integral in equation (A.4.1), as well as $1/T_\infty$, the integral, denoted "I," becomes

$$I = \frac{1}{\sigma^2} \int_{1-a}^{u} \frac{\log(u'+a)}{u'} du' \, ; \, u \equiv \exp(\sigma[\zeta - \zeta_\ell]) - a. \tag{A.4.5}$$

To perform this integral for values of $a > 0.5$, one identifies two regions of the domain of integration, i.e., $u' \leq a$ and $u' > a$, so that, depending on the values of a, $\sigma$, and the desired altitude $\zeta$, the integral I might require up to two separate integrations. If $a < 0.5$, $u \geq 1 - a > a$, for all values of u in the altitude domain of integration ($\zeta'' \in [0, \zeta - \zeta_\ell]$), so that only one integral is necessary.

This example, below, considers a problem in which $a > 0.5$ and the upper limit $u > a$, so that two subdomains (integrals) are required, i.e., $I = I_A + I_B$, corresponding respectively to the integration variable $u' \leq a$ and to $u' > a$. This happens in the upper portion of the transition region, e.g., when $z_\ell \sim 120$ km and $z_u > 250$ km, as is the case for two of the major neutral species (O, $O_2$) in NRLMSISE-00. The lower integral is

$$I_A = \frac{1}{\sigma^2} \int_{1-a}^{a} \frac{\log(u'+a)}{u'} du' = \frac{1}{\sigma^2} \left[ \int_{1-a}^{a} \frac{\log a}{u'} du' + \int_{1-a}^{a} \frac{\log(1+\frac{u'}{a})}{u'} du' \right]. \tag{A.4.6}$$

Denoting this as $I_A = I_{A1} + I_{A2}$, the first integral in brackets is

$$I_{A1} = \frac{1}{\sigma^2} \log a \log u' \Big|_{1-a}^{a} = \frac{1}{\sigma^2} \log a \log \frac{a}{1-a} = \frac{1}{\sigma^2} \log\left(\frac{T_\infty - T_\ell}{T_\infty}\right) \log\left(\frac{T_\infty - T_\ell}{T_\ell}\right). \quad (A.4.7)$$

To compute the second integral in brackets, expand the logarithm as a series:

$$I_{A2} = \frac{1}{\sigma^2} \int_{1-a}^{a} \frac{1}{u'} \left(\frac{u'}{a} - \frac{u'^2}{2a^2} + \frac{u'^3}{3a^3} - \frac{u'^4}{4a^4} + \ldots\right) du' = -\frac{1}{\sigma^2} \sum_{n=1}^{\infty} \frac{(-1)^n}{n^2} \left(\frac{u'}{a}\right)^n \Big|_{1-a}^{a}$$

$$= -\frac{1}{\sigma^2} \sum_{n=1}^{\infty} \frac{(-1)^n}{n^2} \left(1 - \left\{\frac{1-a}{a}\right\}^n\right) = -\frac{1}{\sigma^2} \sum_{n=1}^{\infty} \frac{(-1)^n}{n^2} \left(1 - \left\{\frac{T_\ell}{T_\infty - T_\ell}\right\}^n\right). \quad (A.4.8)$$

The upper integral is

$$I_B = \frac{1}{\sigma^2} \int_a^u \frac{\log(u'+a)}{u'} du' = \frac{1}{\sigma^2} \left[\int_a^u \frac{\log u'}{u'} du' + \int_a^u \frac{\log(1+\frac{a}{u'})}{u'} du'\right]. \quad (A.4.9)$$

Denoting this as $I_B = I_{B1} + I_{B2}$, the first integral in brackets, with a change of variable to $v = \log u'$, is

$$I_{B1} = \frac{1}{\sigma^2} \frac{v^2}{2} \Big|_{\log a}^{\log u} = \frac{1}{2\sigma^2} \left[(\log u)^2 - (\log a)^2\right]. \quad (A.4.10)$$

To compute the second integral in brackets, expand the logarithm as a series:

$$I_{B2} = \frac{1}{\sigma^2} \int_a^u \frac{1}{u'} \left(\frac{a}{u'} - \frac{a^2}{2u'^2} + \frac{a^3}{3u'^3} - \frac{a^4}{4u'^4} + \ldots\right) du' = \frac{1}{\sigma^2} \sum_{n=1}^{\infty} \frac{(-1)^n}{n^2} \left(\frac{a}{u'}\right)^n \Big|_a^u$$

$$= \frac{1}{\sigma^2} \sum_{n=1}^{\infty} \frac{(-1)^n}{n^2} \left(\left\{\frac{a}{u}\right\}^n - 1\right) = \frac{1}{\sigma^2} \sum_{n=1}^{\infty} \frac{(-1)^n}{n^2} \left(\left\{\frac{T_\infty - T}{T}\right\}^n - 1\right). \quad (A.4.11)$$

The expression uses equation (A.2.1). Note that the corresponding nonparametric terms involving unity in the final expressions for $I_{A2}$ and $I_{B2}$ reinforce each other in I rather than canceling. This is because different factors (log a and log u', respectively) were necessarily factored from the integrands to allow expansion of log (u'+a) in the respective variables u'/a and a/u'. These variables are mutually inverse and are necessarily less than unity to facilitate the expansions in $I_{A2}$ and $I_{B2}$.

With these integrals, equation (A.4.1) becomes

$$\log n_{i2}(\zeta) = -\frac{g_0}{k_B T_\infty}\frac{m_u - m_\ell}{\zeta_u - \zeta_\ell}\left\{\begin{array}{l}\dfrac{1}{\sigma^2}\log a \log\dfrac{a}{1-a} - \dfrac{1}{\sigma^2}\sum_{n=1}^{\infty}\dfrac{(-1)^n}{n^2}\left(1 - \left\{\dfrac{1-a}{a}\right\}^n\right) \\ + \dfrac{1}{2\sigma^2}\left[(\log u)^2 - (\log a)^2\right] + \dfrac{1}{\sigma^2}\sum_{n=1}^{\infty}\dfrac{(-1)^n}{n^2}\left(\left\{\dfrac{a}{u}\right\}^n - 1\right)\end{array}\right\} \quad \text{.(A.4.12)}$$

Alternatively, factoring out $1/\sigma^2$, combining the first and third terms in brackets, and expressing u and a in terms of temperature and temperature parameters convert the expression to

$$\log n_{i2}(\zeta) = -\frac{g_0}{k_B T_\infty}\frac{1}{\sigma^2}\frac{m_u - m_\ell}{\zeta_u - \zeta_\ell}\left\{\begin{array}{l}\dfrac{1}{2}\left[\left(\log\dfrac{T}{T_\infty}\dfrac{T_\infty - T_\ell}{T_\infty - T}\right)^2 + \left(\log\dfrac{T_\infty - T_\ell}{T_\infty}\right)^2 - 2\log\left(\dfrac{T_\infty - T_\ell}{T_\infty}\right)\log\dfrac{T_\ell}{T_\infty}\right] \\ -\sum_{n=1}^{\infty}\dfrac{(-1)^n}{n^2}\left(1 - \left\{\dfrac{T_\ell}{T_\infty - T_\ell}\right\}^n\right) + \sum_{n=1}^{\infty}\dfrac{(-1)^n}{n^2}\left(\left\{\dfrac{T_\infty - T}{T}\right\}^n - 1\right)\end{array}\right\}$$

.(A.4.13)

Note again that in this case, a > 0.5, so that $T_\ell < 0.5\, T_\infty$, and that the upper limit of the integral (A.4.5) is u > a so that $T > 0.5\, T_\infty$. In the last expression of the equation, the first term in brackets dominates as T increases and diverges as T approaches $T_\infty$. The leading negative sign on the RHS of the equation ensures that the species density decreases in that limit. The two series each converge to an expression less than unity in magnitude. While either series might converge slowly, depending on proximity of the argument in parentheses to unity, the algorithm of Ginsberg and Zaborowski [1975] removes this limitation.

**A.4.2. Alternative Solution and General Polynomial Interpolation**

While technically a closed−form solution, equations (A.4.12−13) are inelegant, computationally inconvenient, and physically unrevealing at best. Further, a more general polynomial interpolation will probably be more difficult to integrate and with a more complex result. Fortunately an alternative approach seems more promising in these aspects.

Returning to equation (A.4.1), one sees that the denominator of the integrand has the form $1 − \alpha$ where $\alpha < 1$. Expanding $1/(1 − \alpha)$ as a convergent geometric series in $\alpha$ and integrating term by term in the variable, $\zeta'' \equiv \zeta' - \zeta_\ell$, over the domain $[0, \zeta - \zeta_\ell]$ gives us

$$\log n_{i2}(\zeta) = -\frac{g_0}{k_B T_\infty}\frac{m_u - m_\ell}{\zeta_u - \zeta_\ell}\left[\frac{(\Delta\zeta)^2}{2} + \frac{1}{\sigma^2}\sum_{n=1}^{\infty}\left\{\frac{a^n}{n^2}\left(1 - e^{-n\sigma(\Delta\zeta)}[1 + n\sigma(\Delta\zeta)]\right)\right\}\right], \quad (A.4.14)$$

where $\Delta\zeta \equiv \zeta - \zeta_\ell$. Note that this requires the evaluation of one exponential function and several additions and multiplies, depending on convergence criteria. Regardless of the criterion, this is easy to program and compute. A sample calculation at $\Delta\zeta = 1/\sigma = 50$ km indicates that the fifth term contributes less than 1% of the total value.

In the case of a higher order polynomial interpolation, the same method can generate an expression similar to equation (A.4.14) for each order of the interpolant. For each order above 1, the last square bracket will contain a polynomial having terms of alternating sign. The point is that convergence continues to be assured, calculation is relatively simple, and for each order, only a moderate number of terms of the sum will be necessary to achieve the desired convergence.

Equation (A.4.14) is easy to test and explore via straightforward investigation of various limits and altitude regimes within the transition region. For example, consider small values of $\Delta\zeta$ relative to the temperature scale height $\sigma$. Expanding the exponential through second order gives us the result

$$\log n_{i2}(\zeta) \approx -\frac{g_0}{k_B T_\infty} \frac{m_u - m_\ell}{\zeta_u - \zeta_\ell} \left[ \frac{(\Delta\zeta)^2}{2} \frac{1}{1-a} \right], \quad (A.4.15)$$

consistent with inspection of equation (A.4.1).

### Appendix B. Relationship of Interspecies Momentum Transfer to Diffusion

This appendix briefly explores the relationship, equation (2.4), between the interspecies drag frequency $\omega_{ij}$ and the diffusion coefficient, $D_{ij}$ suggested by CC58 (Note I) and, in more detail, CC70 (Sections 6.62−6.63). The diffusion equation of CC70 is

$$\mathbf{d}_i + k_{Ti} \nabla \ln T = \sum_j \frac{n_j n_i}{n^2 D_{ij}} (\mathbf{v}_j - \mathbf{v}_i), \quad (5.1.1)$$

where

$$d_i = \frac{p_i}{p} \left\{ \frac{1}{\rho_i} \frac{d}{dz} p_i + g(z) - \frac{1}{\rho} \frac{d}{dz} p - g(z) \right\}, \quad (B.1)$$

from equation (5.1.2) for altitude variation only.

Substituting equations (2.1) and (2.6) into (B.1) gives us

$$d_i = \frac{p_i}{p} \left\{ \left( \frac{d\mathbf{v}_0}{dt} - \frac{d\mathbf{v}_i}{dt} \right) + \frac{1}{\rho_i} \sum_j \omega_{ij} (\mathbf{v}_j - \mathbf{v}_i) \right\}, \quad (B.2)$$

Equation (5.1.1) becomes

$$\frac{p_i}{p} \left\{ \left( \frac{d\mathbf{v}_0}{dt} - \frac{d\mathbf{v}_i}{dt} \right) + \frac{1}{\rho_i} \sum_j \omega_{ij} (\mathbf{v}_j - \mathbf{v}_i) \right\} + k_{Ti} \frac{d \ln T}{dz} = \sum_j \frac{n_j n_i}{n^2 D_{ij}} (\mathbf{v}_j - \mathbf{v}_i), \quad (B.3)$$

Substituting the CC approximation of $\omega_{i,j}$ in equation (2.4) results in

$$\frac{p_i}{p} \left( \frac{d\mathbf{v}_0}{dt} - \frac{d\mathbf{v}_i}{dt} \right) + \sum_j \frac{p_i p_j}{p^2 D_{ij}} (\mathbf{v}_j - \mathbf{v}_i) + k_{Ti} \frac{d \ln T}{dz} = \sum_j \frac{n_j n_i}{n^2 D_{ij}} (\mathbf{v}_j - \mathbf{v}_i) \quad (B.3)$$

The second terms on the two sides cancel, leaving us to conclude that the difference between the acceleration of the composite fluid and that of any species is attributable entirely to thermal diffusion, which is itself not a force term! Such a situation is contradictory at best.

Further, if thermal diffusion were to be negligible, the accelerations would be essentially equal, which is the case only for strong coupling in a realistic thermosphere. Indeed CC does favor an approximation of nearly equal accelerations of the composite fluid and each separate component. However, this is not general, especially in the upper thermosphere and transition region; so the approximation of equation (2.4) is not generally applicable to thermospheric dynamics.

The problem might lie with the ambiguity in interpreting the velocities in equation (5.1.1), as discussed in Section 5.1 (paragraph following equation (5.1.2)). Perhaps a better understanding of this situation will emerge with the use of higher order terms in various expansions of the Boltzmann distribution functions. Until that is explored, caution is advisable when using equation (2.4).

**Acknowledgements**

The authors appreciate financial support for this work, as provided by NASA, the Chief of Naval Research, and the Civil Service Retirement System (Picone), along with logistical support by the Naval Research Laboratory (NRL) Volunteer Emeritus Participant Program (Picone). We also greatly appreciate many helpful discussions with NRL colleagues Dr. Judith Lean, Dr. Robert Meier, and Dr. Fabrizio Sassi.